# In-flight Calibration of the Magnetometer on the Mars Orbiter of Tianwen-1

Zou, Zhuxuan[1,2], Yuming Wang[1,2,*], Tielong Zhang[2,3], Guoqiang Wang[4], Sudong Xiao[4], Zonghao Pan[1,2], Zhoubin Zhang[5], Wei Yan[5], Yang Du[6], Yutian Chi[7], Long Cheng[1,2], Zhiyong Wu[1,2], Xinjun Hao[1,2], Yiren Li[1,2], Kai Liu[1,2], Manming Chen[1,2], Zhenpeng Su[1,2], Chenglong Shen[1,2], Mengjiao Xu[7], and Jingnan Guo[1,2]

1 School of Earth and Space Sciences/Deep Space Exploration Laboratory, University of Science and Technology of China, Hefei 230026, China

2 CAS Center for Excellence in Comparative Planetology/CAS Key Laboratory of Geospace Environment/Mengcheng National Geophysical Observatory, University of Science and Technology of China, Hefei 230026, China

3 Space Research Institute, Austrian Academy of Sciences, Graz, Austria

4 Institute of Space Science and Applied Technology, Harbin Institute of Technology, Shenzhen, 518055 China

5 National Astronomical Observatories, Chinese Academy of Sciences, Beijing, China

6 Shanghai Institute of Satellite Engineering, Shanghai, China

7 Institute of Deep Space Sciences, Deep Space Exploration Laboratory, Hefei 230026, China

*Corresponding author, Email: ymwang@ustc.edu.cn

# Abstract

Mars Orbiter Magnetometer (MOMAG) is one of seven science payloads onboard Tianwen-1's orbiter. Unlike most of the satellites, Tianwen-1's orbiter is not magnetically cleaned, and the boom where placed the magnetometer's sensors is not long enough. These pose many challenges to the magnetic field data processing. In this paper, we introduce the in-flight calibration process of the Tianwen-1/MOMAG. The magnetic interference from the spacecraft, including spacecraft generated dynamic field and slowly-changing offsets are cleaned in sequence. Then the calibrated magnetic field data are compared with the data from the Mars Atmosphere and Volatile EvolutioN (MAVEN). We find that some physical structures in the solar wind are consistent between the two data sets, and the distributions of the magnetic field strength in the solar wind are very similar. These results suggest that the in-flight calibration of the MOMAG is successful and the MOMAG provides reliable data for scientific research.

Keywords: magnetometer, in-flight calibration, Martian magnetic field, Tianwen-1

## 1. Introduction

Tianwen-1, the first China's Mars exploration mission, contains an orbiter running on a large elliptical polar orbit[1]. It can arrive 2.8-4.3 $R_M$ (Radius of Mars) to observe the farther magnetotail region[2], filling the gap of the existing Mars explorations, e.g., Mars Global Surveyor (MGS)[3] and the Mars Atmosphere and Volatile EvolutioN (MAVEN)[4]. Its magnetometer (MOMAG) is designed to observe the magnetic field around Mars and has three main goals with cooperation with other instruments: (1) to explore the in-situ environment of the Martian ionosphere, induced magnetosphere and interplanetary space, (2) to study the interaction mechanism between these regions, (3) to study the ionospheric conductivity and currents together with the measurements of Mars rover magnetometer[1, 5].

Most magnetometers are placed on the top of a long boom to keep far away from the spacecraft which may have significant interference on the measured magnetic field, or are required a strict magnetic cleanliness programme[6-9]. Venus Express is an exception, of which the spacecraft was not magnetically cleaned, but the magnetometer on board the spacecraft successfully provided excellent data after the careful in-flight calibration[10]. Tianwen-1 orbiter does not equip the magnetic cleanliness program due to budget. The measured magnetic field data could be severely affected by operations of the instruments onboard the orbiter, solar array driving mechanism changes, antenna transmission effects, reaction wheel effects, and slow thermal drift[11], which bring a challenge for data processing. Thus, the in-flight calibration becomes very important for MOMAG to provide reliable scientific data.

The magnetic field in the solar wind has some natural physical characteristics, which could be used in the magnetometer calibration. In the early stage, researchers found that over some solar rotations the averages of the magnetic field components are zero, and the method could only get an offset for several months consequently[12]. Later, some calibration methods based on the Alfvénic fluctuation, which occurs frequently in the solar wind and during which the magnetic field strength remains almost unchanged, were developed[12-16]. Davis-Smith method is a classical method based on the assumption that the variance of the squared magnetic field magnitude is minimum[15]. Other methods have different assumptions such as that the changes in the field magnitude and the changes in the inclination of the field to any one of the three coordinate axes have no correlation[13, 16], and the fluctuations are transverse to the ambient field[16].

Afterwards, some methods based on mirror mode structures and current sheets were also developed for the zero offset calibration[17-20]. They assumed that the magnetic field is parallel to the maximum variance direction and the minimum variance direction, respectively, according to the minimum variance analysis (MVA)[21]. Mirror mode structures and current sheets obviously occur less than the Alfvénic fluctuation events in solar wind. However, they may be necessary complementary for the calibration in the magnetosheath or magnetosphere.

In the first several months of MOMAG routine operation, the orbital period is about 7.8 hr with more than 50% of time in the solar wind. Thus, we can use the properties of Alfvénic fluctuation to calibrate the MOMAG data. Here, the recently developed Wang-Pan method [14] is used. In the next section, we first introduce the clean method of the spacecraft generated dynamic field due to the operation of instruments. In Sect.3 we introduce the calibration of the MOMAG's zero offset. Sect.4 shows the processed data and compares them with the calibrated magnetic field data obtained by the MAVEN spacecraft. Sect.5 briefly summarizes the main results of this paper.

## 2. Clean Method for Dynamic Field

MOMAG has two sensors, one is mounted at the top of the boom (outer sensor), which is 3.19 m away from the orbiter surface, and the other is at the middle of the boom (inner sensor), 2.29 m away from the orbiter surface. Both of them measure the magnetic field independently, which is severely interfered by spacecraft operations. Such interferences contain two types. One appears like an artificial jump in the magnitudes of magnetic field components due to the operations, such as turning on/off, of the instruments onboard the orbiter. It frequently occurs with a rate of hundreds of times a day. The other type is a kind of systematic offset including the offset of the sensors themselves, which is quite stable and does not change too much over hours to days.

Since the dynamic magnetic field originating from the orbiter decreases monotonously with the distance away from the orbiter, the interferences acting on the two sensors are different. Ness et al[22] proposed a theoretic method, called gradiometer technique, to remove spacecraft interference by dual sensors. One of the basic assumptions is that the spacecraft magnetic field could be approximated as a dipole field. Here, we use a similar method. For the first type of the interferences, the artificial jumps, their magnitudes at the two sensors are different. Thus, we can easily recognize them and distinguish them from real jumps in the solar wind. Then we get rid of these artificial jumps by using the algorithm that was successfully applied on the magnetometer of Venus Express[11]. One can refer to that paper for details.

Figure 1a shows the three components of the original magnetic field detected by the two sensors between 01:30 and 04:15 UT on 17 November 2021. Note that the zero offset is not cleaned yet, and therefore the values of the measured magnetic field are meaningless for the scientific research. For comparison purposes, we subtract their mean and translate them to near zero. We can find that the original data have lots of jumps in magnitude. Some of them are physical but more are artificial, which have been marked by purple lines. The artificial jumps at inner sensor are larger than those at outer sensor. After cleaning these artificial jumps, we could find that the magnetic field components become much more stable than original data, as shown in Figure 1b.

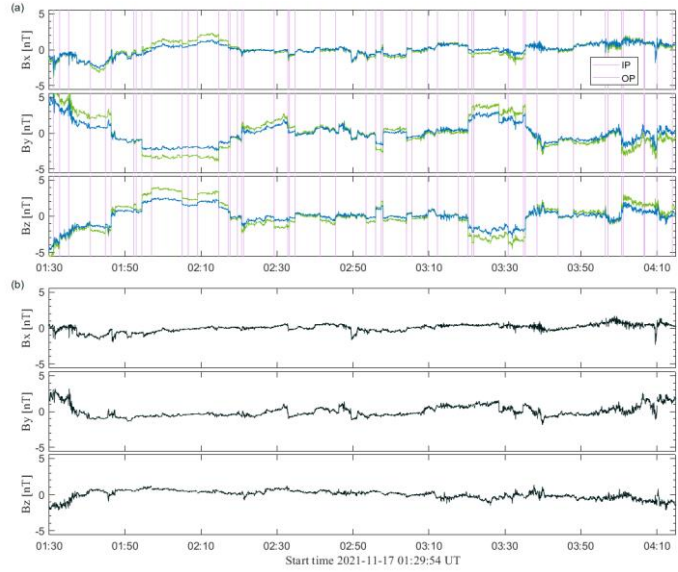

Figure 1. (a) The three components of the original magnetic field detected by the outer probe (blue) and inner probe (green) between 01:30 and 04:15 UT on 17 November 2021. The dynamic fields are marked in purple lines. (b) The magnetic field with the dynamic fields removed (black). The three components are presented in the spacecraft coordinates.

## 3. Correction method for the Zero Offset

After removing the dynamic fields, only the slowly-changing zero offset is unremoved in the data. We use the Alfvénic fluctuation events in the solar wind to calibrate it based on the Wang-Pan method [23]. We first develop a semi-automated procedure to choose highly Alfvénic fluctuation events as follows.

The selection of Alfvénic fluctuation events may affect the precision of the zero offset. Thus, it is a key step to acquire enough highly Alfvénic fluctuation events. In well-calibrated magnetic field data, researchers have some methods such as characterizing Alfvén wave as a fluctuation where the magnitude of magnetic field changes little or utilizing the high correlation between magnetic field vector and the plasma velocity [12]. However, in the uncalibrated data, we only have the magnetic field with a zero offset. Therefore, the constant magnitude of magnetic field with some components changing is the only criterion that an Alfvenic region should meet. The selected candidate Alfvenic fluctuations should be long enough in time to suit the Wang-Pan method but not too long to be mixed with compressional structures. Since the Alfvenic fluctuation events always exist in the quiet solar wind, we first manually and roughly select the time intervals without notable structures in the quiet solar wind. Then we search the candidate Alfvenic fluctuation events in three components of magnetic field, respectively. We use a 5s-lowpass Butterworth filter to get rid of high-frequency noise, and further use a 300s-lowpass Butterworth filter to acquire background magnetic field. Then we search the time interval according to the following criteria: the variation of the magnetic field is greater than 0.6 nT and the magnetic field crosses the background value at least 3 times within 0.2-10 minutes, in the 5s-lowpass data for each magnetic field component. Any time interval meeting the above criteria is selected to be a candidate.

Figure 2 gives a sample period of Alfven fluctuation event selection in the spacecraft coordinates from 02:28 to 03:10 UT on 17 November 2021. The selected Alfvenic candidate intervals in each component are indicated by dark-light alternating shadowed regions in Figure 2a. For each magnetic field component, we can find many candidate Alfvenic intervals. If one candidate Alfvenic interval found in one component overlaps with a candidate interval in another component, we merge them as a single Alfvenic fluctuation event. The start and end time of the component with the larger variance is chosen as the start and end time of the merged Alfvenic event. The merged Alfvenic intervals during this period are shown in Figure 2b.

The next step is to check if the selected merged intervals could be almost constant in the total magnetic field strength, which is the characteristic of a typical Alfvenic wave, after shifting with an offset. We test the possible offset vector in a range of $\pm 10$ nT centering on the mean value of each Alfvenic interval with the step of 0.4 nT, i.e., a total of 50 grid points in each component. The variance of the total magnetic field strength, given by $\frac{\sum_i (B_i - \langle B \rangle)^2}{N}$, in which $B_i$ is the magnetic field strength of each data points after a test offset is applied during the interval, $\langle B \rangle$ is the averaged value of the magnetic field strength during the interval and $N$ is the total number of the data points during the interval, is calculated to evaluate how well the criterion of the constant magnitude of an Alfven wave is satisfied. In our procedure, the variance should less than 0.012 nT$^2$. After testing all the possible offsets, we get the distribution of the variance in the searching cube. All the candidate intervals satisfy the requirements of the variance are treated as the finally selected Alfvenic fluctuation event.

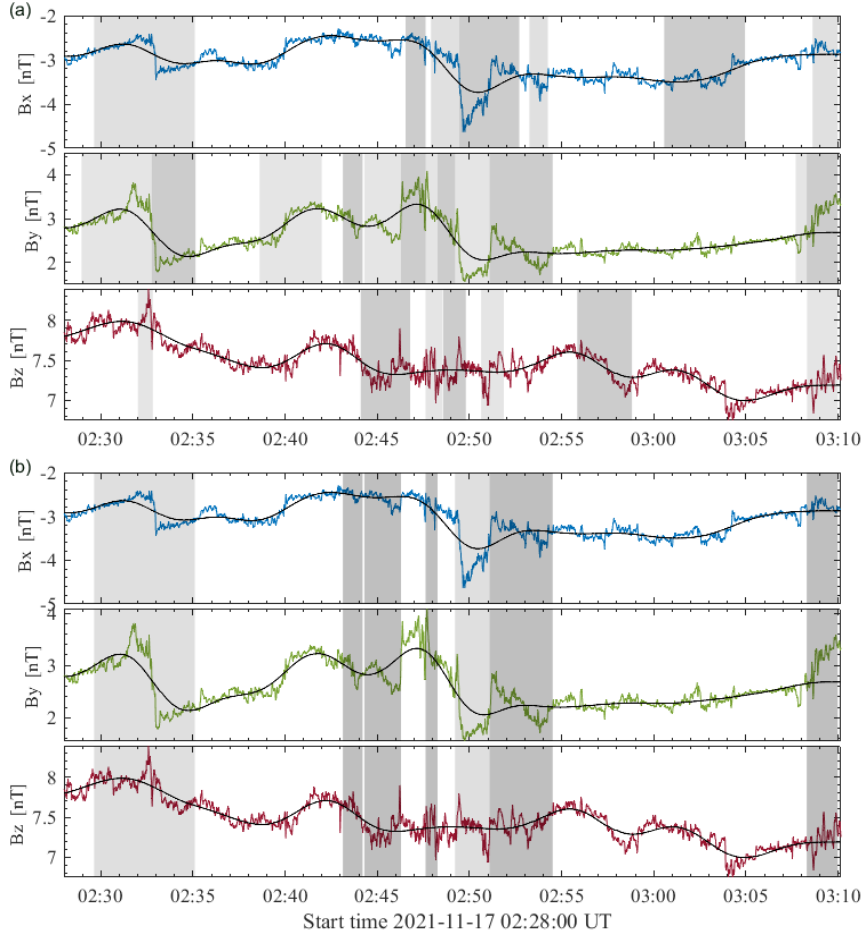

Figure 2. Examples of finding Alfven fluctuation events on 17 November 2021. Colored lines are the partially calibrated data and black lines are their ambient magnetic field. (a) The candidate Alfvenic intervals in each magnetic field component in grey shades. (b) The merged candidate Alfvenic fluctuation events are in gray areas, and the dark gray areas are finally used.

For each finally selected Alfvenic fluctuation event, we investigate all the slices perpendicular to one direction, e.g., to the z-axis as shown in Figure 3a, and locate the position of the minimum variance in each slice. Then we fit these positions with a linear line as indicated by the blue line in Figure 3a. We repeat the above procedure for the other two directions, i.e., the x and y-axes, and choose the best fit one as the optimal offset line (OOL) of the Alfvenic fluctuation event. It means that we could get an OOL for each Alfvenic fluctuation event. Then for every 10 adjacent Alfvenic fluctuation events, i.e., for every 10 OOLs, we locate a point, i.e., the position of the most optimal offset, from which the sum of the distances to all the 10 OOLs is the shortest as illustrated by Figure 3b. Note, since the qualities of the finally selected Alfvenic fluctuation events are different, we assign the OOLs different weights following the method in the paper by Hu et al [23], and therefore the distances to the 10 OOLs have different weights when we calculate the sum of the distances.

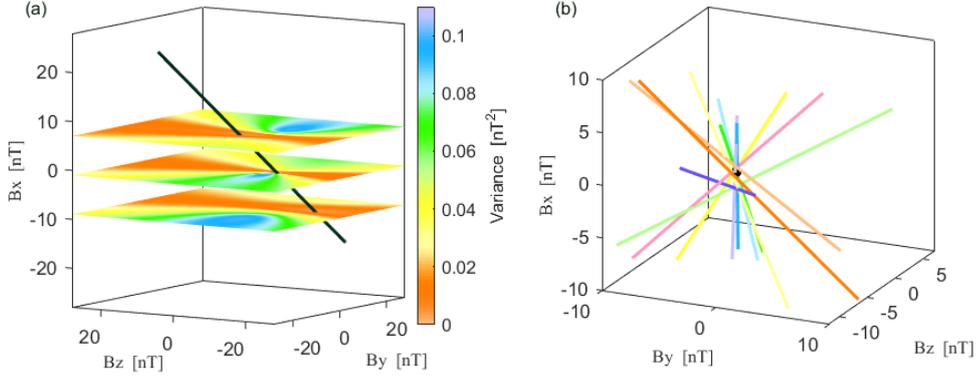

Figure 3. A diagram of solving offset process. (a) The solution procedure of the OOL. The colored slices in the offset cube represent the variances of an Alfven interval added the test offsets in the plane. The points with the minimum variance in each slice are fitted into the black line called OOL. (b) A zero offset (black point) formed by ten optimal offset lines shown in colored lines.

During the period from November 16 to December 31, 2021, we totally find 824 Alfvenic fluctuation events corresponding to 824 OOLs. Figure 4 displays the obtained offsets for the outer sensor during the period of interest with the orange dots. Considering that the offset should slowly change with time, we smooth and extrapolate the scattered offsets to a relatively smooth curve. We smooth the offset in a 72 hours' sliding window, which is nearly nine orbit periods, by using the local regression with weighted linear least squares and a 2nd degree polynomial model. Then we use the linear interpolation method to get the offset values throughout the period. It means that the offsets obtained based on the Alfvenic fluctuation events in solar wind are also applied to the magnetosheath region.

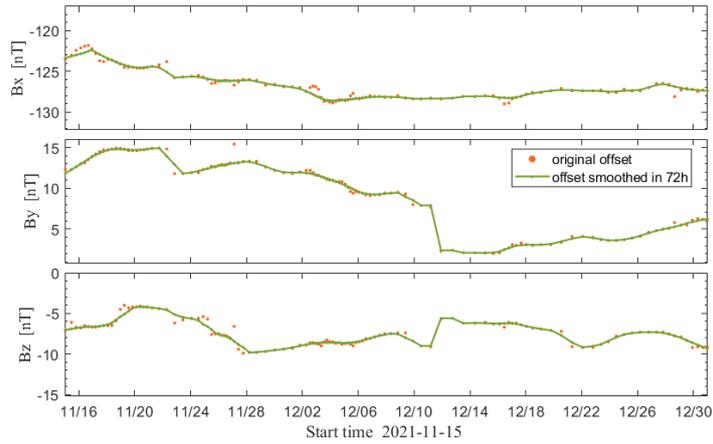

Figure 4. The calculated offset (orange) and the smoothed offset (green) in spacecraft coordinate system during November 16 and December 31, 2021.

## 4. The Calibrated Data

After removing the artificial jumps and offsets, we obtain the scientific Level 2 (or called Level C following China's convention) data of MOMAG by converting the magnetic field vectors from spacecraft coordinates into physical-based coordinates, e.g., the Mars-centered Solar Orbital (MSO) coordinates. Figure 5a shows the processed data between 12:50 and 20:30 UT on 26 November 2021 in blue lines. The orbit is shown in Figure 5b in the color-coded line. From the figure, we can see that the orbiter crossed the Martian bow shock and entered the magnetosheath at ~15:34 UT, and crossed the bow shock again at ~18:20 UT when it flew from the magnetosheath to the solar wind.

Thank to MAVEN still working at Mars, we may assess the reliability of the calibrated data by comparing the data with the MAVEN's MAG data[4] shown in green lines in Figure 5a. The orbital period of Maven is ~4 hr and it came into the magnetosphere twice during the period of interest. We could compare the magnetic field detected by them when they were both in the solar wind, which are denoted by shadows. The magnetic field they detected are similar in the solar wind. The mean magnetic field of Tianwen-1 in the solar wind in x, y, z direction in MSO coordinate system were -0.6 nT, 0.5 nT and 0.9 nT respectively. In MAVEN's observation, they were -0.6 nT, 1.7 nT and 0.9 nT. The mean magnitudes were 3.2 nT in Tianwen-1 and 3.6 nT in MAVEN. At ~20:14 UT both the spacecraft observed a sudden change. In the x component, the magnetic field at both spacecraft jumped from negative to positive, while in the other two components, the magnetic field from positive to negative with slightly difference in magnitude. Tianwen-1 orbiter has a quite different orbit from MAVEN as illustrated in Figure 5b. The distance between Tianwen-1 orbiter and MAVEN was more than thousands of kilometers at that time. Thus, it is natural that there are differences between the data from the two spacecrafts.

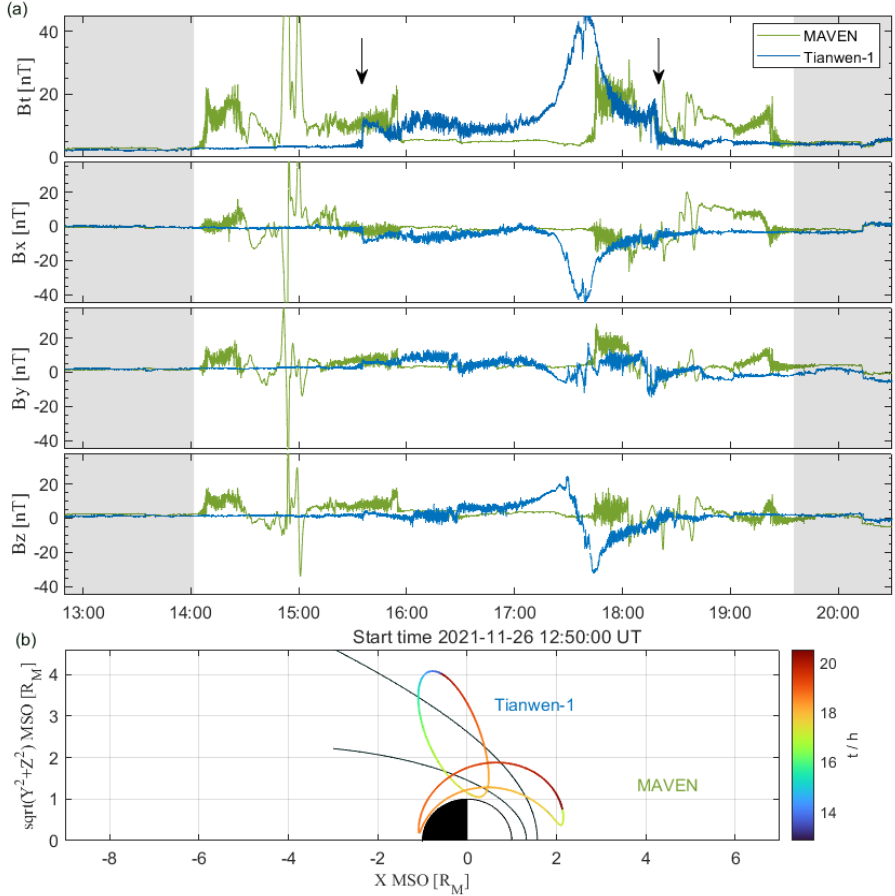

Figure 5. The scientific data detected by Tianwen-1 and MAVEN between 12:50 and 20:30 UT on 26 November 2021. (a) The processed magnetic field in MSO coordinate system. The arrows indicate the positions of the bow shock. The periods when the two satellites are both in the solar wind are marked in shadows. (b) Their orbit position in colored line.

Figure 6 shows the processed data and the corresponding MAVEN magnetic field between 05:00 and 13:00 UT on 07 December 2021 in the same format as Figure 5. The magnitude and components of their magnetic field are similar in the solar wind. The mean magnetic field of Tianwen-1 in x, y, z direction in the solar wind were 0.4 nT, -0.4 nT and -0.6 nT, respectively. In MAVEN observation, they were 0.9 nT, -0.4 nT and -0.2 nT. Both of the mean magnitudes were 1.3 nT. Different from the bow shocks in Figure 5, the signature of the bow shock observed by Tianwen-1 and MAVEN is not clear because of the upstream solar wind conditions. Besides, around 12:31 UT, both MAVEN and Tianwen-1 observed a set of strong fluctuations of the magnetic field.

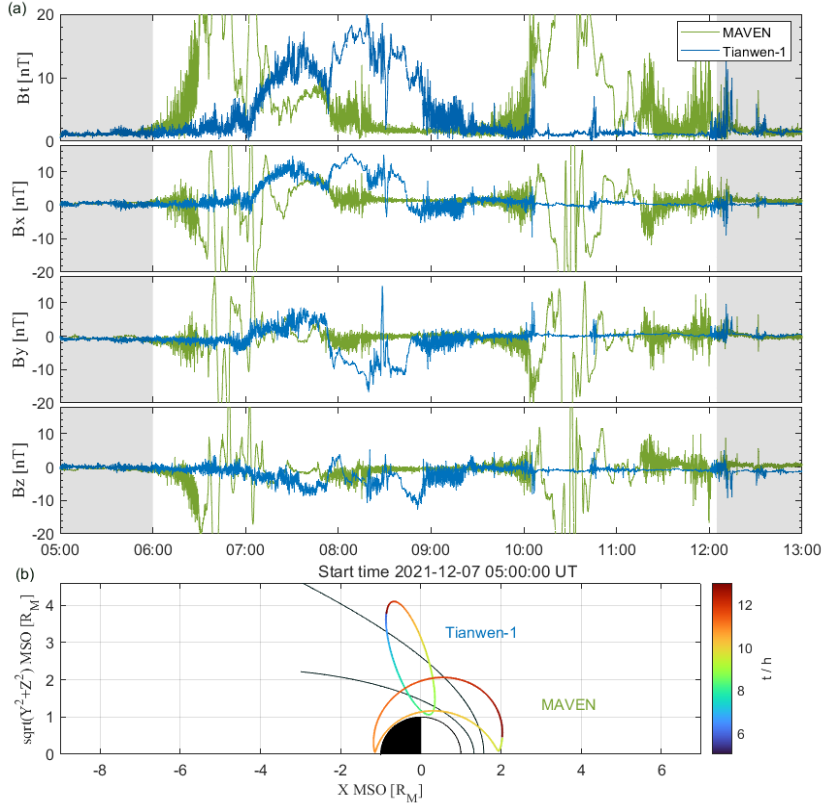

Figure 6. The scientific data detected by Tianwen-1 and MAVEN between 05:00 and 13:00 UT on 07 December 2021 in the same format as Figure 5.

Except the featured structures we compared in the above two episodes, we also investigate some magnetic holes (MHs) that are widely detected in the space environments[24], especially in the solar winds[25, 26]. Some studies have detected the MHs in the solar wind surrounding Mars[27]. The typical characteristic of a MH is that the magnetic field amplitude has a large dip. An example is given in Figure 7. There was a serial of MHs in just five minutes on 17 November 2021. The magnetic field magnitude shown in Figure 7a have some large dips, which are corresponding to the MHs. Figure 7b shows the three magnetic field components in MSO coordinates. The directions of their magnetic field changed slowly, and therefore they could be called linear MHs.

We rotate the magnetic field to the LMN coordinate system shown in Figure 7c using the MVA method. Because we apply the MVA method in over 5 minutes duration rather than the duration of only one MH, the magnetic field in the maximum variation directions ($B_L$) denotes the slow increasing of the ambient magnetic field. The characteristic of large dips appears in the intermediate variation directions (M), and the ambient magnetic field is mainly in the L direction.

We also find a series of MH structures around 17:04 UT in MAVEN/MAG data shown in Figure 7d-7f. Their dips are smaller than the MHs observed by the Tianwen-1. They appeared earlier at MAVEN than at Tianwen-1, since MAVEN was at the upstream of Tianwen-1. Whether they are mirror modes and their creation environment require further confirmation.

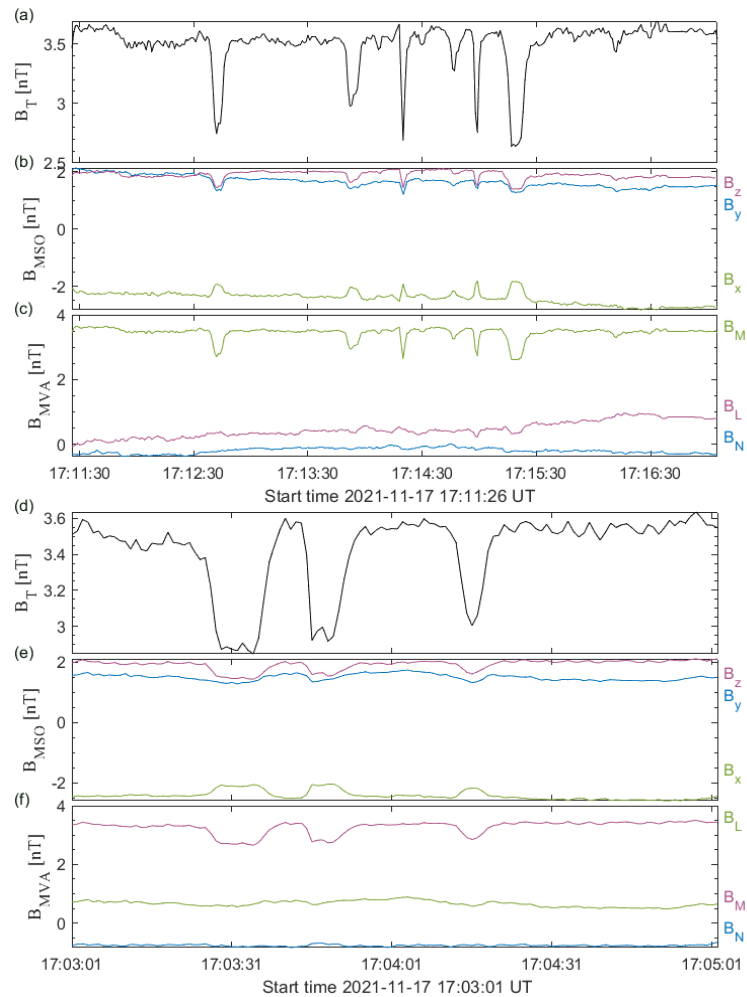

Figure 7. The serial magnetic holes observed in Tianwen-1 and Maven on 17 November 2021. (a) The magnetic field magnitude of Tianwen-1. (b) Three magnetic field components of Tianwen-1 in MSO coordinates. (c) Three magnetic field components of Tianwen-1 in LMN coordinates. (d) The magnetic field magnitude of MAVEN. (e) Three magnetic field components of MAVEN in MSO coordinates. (f) Three magnetic field components of MAVEN in LMN coordinates.

Alternatively, we can statistically compare the distributions of the total magnetic field strength in the solar wind from the two spacecraft as shown in Figure 8. Based on the data during November 16 and December 31, 2021, there are about 314 hr MAVEN data and 400 hr Tianwen-1 data in the solar wind, and about 117 hr when both MAVEN and Tianwen-1's orbiter stayed in the solar wind. Based on the 117 hr data, it could be found that the magnetic field magnitude distribution from MOMAG is similar to that from MAVEN/MAG. The mean value is 3.02 nT for Tianwen-1 and 3.07 nT for MAVEN. The deviation of the median values is only 0.05 nT, which is negligible compared to the magnetic field strength in the solar wind. The comparison suggests that the measured magnetic fields by the two spacecraft are quite consistent, and our calibrated data is credible.

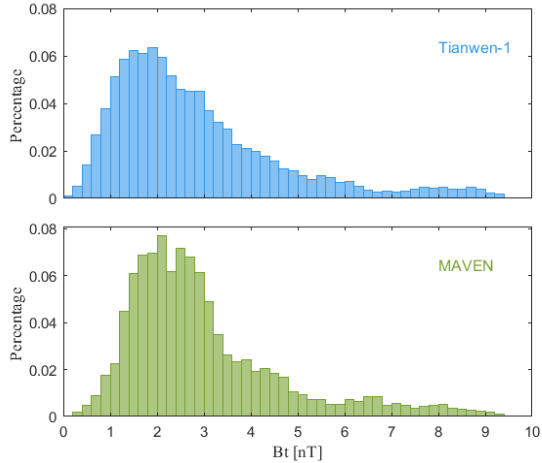

Figure 8. The distributions of the magnetic field strength in the solar wind based on Tianwen-1/MOMAG and MAVEN during November 16 and December 31, 2021.

## 5. Conclusion

In this work, we present the process of the in-flight calibration of the Tianwen-1/MOMAG magnetic field data as well as the data quality. We develop a series of procedures to clean the interference caused by the spacecraft. The dynamic fields are cleaned firstly based on the data from the dual sensors. Then we use properties of Alfvenic waves to correct the offset of measured magnetic field. One and a half months' calibrated magnetic field data are analyzed and compared with MAVEN/MAG data. The common typical structures, such as physical jumps and MHs, could be found in the data from the both spacecrafts. The distribution of magnetic field strength in the solar wind measured by Tianwen-1/MOMAG is similar to that by MAVEN/MAG with only 0.09 nT deviation in the median value. These results suggest that the difference of the data from the two spacecraft could be neglected and the calibrated data are reliable.

**Acknowledgements**

This work in China was supported by the NSFC (Grant Nos 42130204, 42188101 and 42241155) and the Strategic Priority Program of the Chinese Academy of Sciences (Grant No. XDB41000000). Y.W. is particularly grateful to the support of the Tencent Foundation.

**Data Availability**

We acknowledge the use of the data from the MAG and SWI onboard MAVEN spacecraft, which are obtained from NASA Planetary Data System (https://pds-ppi.igpp.ucla.edu/).

The Tianwen-1/MOMAG data are publicly available at CNSA Data Release System (http://202.106.152.98:8081/marsdata/) or the data used in this paper can just be downloaded from the official website of the MOMAG team (http://space.ustc.edu.cn/dreams/tw1_momag/).

**References:**

[1] Zou Y, Zhu Y, Bai Y et al. Scientific Objectives and Payloads of Tianwen-1, China's First Mars Exploration Mission. Adv Space Res, 2021, 2: 812-823

[2] Li C, Zhang R, Yu D et al. China's Mars Exploration Mission and Science Investigation. Space Sci Rev, 2021, 4: 57

[3] Acuna M H, Connerney J, Wasilewski P et al. Magnetic Field and Plasma Observations at Mars: Initial Results of the Mars Global Surveyor Mission. Science, 1998, 5357: 1676-1680

[4] Jakosky B M, Lin R P, Grebowsky J M et al. The Mars Atmosphere and Volatile Evolution (Maven) Mission. Space Sci Rev, 2015, 1-4: 3-48

[5] Liu K, Hao X, Li Y et al. Mars Orbiter Magnetometer of China's First Mars Mission Tianwen-1. Earth Planet Phys, 2020, 4: 384-389

[6] Balogh A. Planetary Magnetic Field Measurements: Missions and Instrumentation. Space Sci Rev, 2010, 1-4: 23-97

[7] Bennett J S, Vyhnalek B E, Greenall H et al. Precision Magnetometers for Aerospace Applications: A Review. Sensors-Basel, 2021, 16: 5568

[8] Dougherty M K, Kellock S, Southwood D J et al. The Cassini Magnetic Field Investigation. Space Sci Rev, 2004, 1-4: 331-383

[9] Balogh A, Dunlop M W, Cowley S et al. The Cluster Magnetic Field Investigation. Space Sci Rev, 1997, 1-2: 65-91

[10] Zhang T L, Baumjohann W, Delva M et al. Magnetic Field Investigation of the Venus Plasma Environment: Expected New Results From Venus Express. Planet Space Sci, 2006, 13-14: 1336-1343

[11] Zhang T L, Delva M, Baumjohann W et al. Initial Venus Express Magnetic Field Observations of

the Venus Bow Shock Location at Solar Minimum. Planet Space Sci, 2008, 6: 785-789

[12] Leinweber H K, Russell C T, Torkar K et al. An Advanced Approach to Finding Magnetometer Zero Levels in the Interplanetary Magnetic Field. Measurement science & technology, 2008, 5: 55104

[13] Belcher J W. A Variation of the Davis-Smith Method for in-Flight Determination of Spacecraft Magnetic Fields. Journal of Geophysical Research, 1973, 28: 6480-6490

[14] Wang G, Pan Z. A New Method to Calculate the Fluxgate Magnetometer Offset in the Interplanetary Magnetic Field: 1. Using Alfvén Waves. Journal of Geophysical Research: Space Physics, 2021, 4: 28893

[15] Davis L, Smith E J. The in-Flight Determination of Spacecraft Magnetic Field Zeros. EOS Transaction American Geophysical Union, 1968, 49:

[16] Hedgecock P C. a Correlation Technique for Magnetometer Zero Level Determination. Space Science Instrumentation, 1975, 1: 83-90

[17] Wang G, Pan Z. A New Method to Calculate the Fluxgate Magnetometer Offset in the Interplanetary Magnetic Field: 2. Using Mirror Mode Structures. Journal of Geophysical Research: Space Physics, 2021, 9: 29781

[18] Plaschke F, Goetz C, Volwerk M et al. Fluxgate Magnetometer Offset Vector Determination by the 3D Mirror Mode Method. Mon Not R Astron Soc, 2017, Suppl_2: S675-S684

[19] Plaschke F, Narita Y. On Determining Fluxgate Magnetometer Spin Axis Offsets From Mirror Mode Observations. Ann Geophys-Germany, 2016, 9: 759-766

[20] Wang G, Pan Z. Fluxgate Magnetometer Offset Vector Determination Using Current Sheets in the Solar Wind. The Astrophysical Journal, 2022, 1: 12

[21] Sonnerup B U Ö, Scheible M. Minimum and Maximum Variance Analysis. ISSI Scientific Report Series, 1998, 185-220

[22] Ness N F, Behannon K W, Lepping R P et al. Use of 2 Magnetometers for Magnetic Field Measurements On a Spacecraft. JOURNAL OF GEOPHYSICAL RESEARCH, 1971, 16: 3564

[23] Hu X, Wang G, Pan Z. Automatic Calculation of the Magnetometer Zero Offset Using the Interplanetary Magnetic Field Based On the Wang-Pan Method. Earth Planet Phys, 2022, 52

[24] Sun W J, Shi Q Q, Fu S Y et al. Cluster and Tc-1 Observation of Magnetic Holes in the Plasma Sheet. Ann Geophys-Germany, 2012, 3: 583-595

[25] Winterhalter D, Smith E J, Neugebauer M et al. The Latitudinal Distribution of Solar Wind Magnetic Holes. Geophys Res Lett, 2000, 11: 1615-1618

[26] Turner J M, Burlaga L F, Ness N F et al. Magnetic Holes in the Solar Wind. Journal of Geophysical Research, 1977, 13: 1921

[27] Madanian H, Halekas J S, Mazelle C X et al. Magnetic Holes Upstream of the Martian Bow Shock: Maven Observations. Journal of Geophysical Research: Space Physics, 2020, 1: 27198